\documentclass[twocolumn]{jpsj3}
\usepackage{txfonts}
\usepackage{amsmath,amssymb,mathrsfs}
\usepackage{graphicx}
\usepackage{overpic}       
\graphicspath{{./fig/}}
\usepackage{wrapfig}
\usepackage{color}

\def\gsim{\; $\raise0.3ex\hbox{$>$}\llap{\lower0.8ex\hbox{$\sim$}}$\;}
\def\lsim{\; $\raise0.3ex\hbox{$<$}\llap{\lower0.8ex\hbox{$\sim$}}$\;}

\title{Snapshot Spectrum  and Critical Phenomenon for Two-Dimensional Classical Spin Systems}

\author{Yukinari Imura$^1$, Tsuyoshi Okubo$^2$, Satoshi Morita$^2$ and  Kouichi Okunishi$^3$}
\inst{$^1$Graduate School of Science and Technology, Niigata University, Niigata 950-2181, Japan \\
$^2$ISSP, University of Tokyo, Kashiwa, Chiba 277-8581, Japan\\
$^3$Department of Physics, Niigata University, Niigata 950-2181, Japan}

\abst{
We investigate the eigenvalue distribution of the snapshot density matrix (SDM) generated by Monte Carlo simulation for two-dimensional classical spin systems.
We find that the distribution in the high-temperature limit is well explained by the random-matrix theory, while that in the low-temperature limit can be characterized by the zero-eigenvalue condensation.
At the critical point, we obtain the power-law distribution with a nontrivial exponent $\alpha\equiv(2-\eta)/(1-\eta)$ and the asymptotic form of the snapshot entropy, on the basis of the relationship of the SDM with the correlation function matrix. 
The aspect-ratio dependence of the SDM spectrum is also mentioned.
}
\begin{document}
\maketitle

\section{Introduction}

The role of the entanglement in quantum spin systems has attracted  much attention, in accordance with recent developments of quantum information physics.
In particular, the entanglement entropy often provides essential information that cannot be accessed by the analysis of the usual bulk physical quantity.\cite{entanglement} 
Moreover, for classical spin models, the well-established correspondence between  $d$-dimensional (D) quantum systems and  $(d+1)$-D classical systems\cite{SuzukiTrotter} enables us to analyze the classical-system version of the entanglement.
According to the path integral representation, the maximum eigenvalue-eigenvector of the transfer matrix for a $(d+1)$-D classical system  basically involves the equivalent implication to the wavefunction of the corresponding $d$-D quantum system. 

For classical spin systems,  the Monte Carlo (MC) simulation has been one of the most powerful numerical approaches to analyzing phase transitions, where the finite-size scaling analysis of expectation values of physical quantities is very effective.
However, the MC sampling has a difficulty in the analysis of the entanglement, which directly requires the wavefunction rather than  expectation values of observables. 
This situation is basically the same for the  quantum MC simulation, except for the Renyi entropy of $n=2$, for which the valence-bond-solid picture of the spin singlet is available.\cite{Sandvikvbs}
Therefore, it is interesting to discuss a possible quantity analogous to the entanglement spectrum/entropy, which is easy to compute by a MC simulation.

In a MC simulation,  snapshots representing the typical spin configurations at the equilibrium are generated, and the average of physical quantities is taken for them. 
An interesting viewpoint is that, although the number of snapshots is huge, each snapshot for a discretized spin model can be regarded as just a bitmap image.
In the field of computer science, moreover, the singular value decomposition (SVD) of the bitmap data is successfully used for the purpose of image compression, where the SVD spectrum characterizes a hierarchical structure embedded in the image.\cite{compress}
Thus, for the MC simulation of the spin system, it is also expected that the eigenvalue spectrum of a reduced snapshot density matrix [See Eqs. (3) and (4)], which we call {\it snapshot spectrum}, reflects essential features associated with the phase transition.
Indeed,  Matsueda has recently conjectured that the snapshot spectrum for the 2D classical spin model might exhibit a similar behavior to the entanglement spectrum for the 1D quantum system.\cite{Matsueda}
In particular, the snapshot entropy at the critical point would have a logarithmic dependence with respect to the cutoff dimension $\chi$, which is reminiscent of the entanglement entropy for the corresponding quantum system\cite{2pscaling}.
However, the snapshot is just a  sample of a typical equilibrium spin configuration, which does not contain the total information equivalent to the wavefunction or the maximal eigenvector of the transfer matrix.
Thus, it is important to thoroughly understand the theoretical background behind such behavior of the snapshot spectrum and entropy.

In this paper, we first investigate the distribution of the snapshot spectrum for the 2D Ising model in detail.
The high-temperature limit is described by the random matrix theory (RMT)\cite{RMT}, whereas the zero-eigenvalue condensation occurs in the low-temperature phase.
At the critical temperature, we find that the snapshot spectrum exhibits a power-law distribution with a nontrivial exponent, which can be explained in connection with the correlation function matrix.
Moreover, we will derive the correct asymptotic form for the $\chi$ dependence of the snapshot entropy.
We also mention the aspect-ratio dependence of snapshot spectrum and numerical results for the 3-state Potts model.

This paper is organized as follows.
In the next section, we explain  definitions of the model and snapshot spectrum.
In section 3, we present results for the square-lattice Ising model in detail.
We also mention  the aspect-ratio dependence of the snapshot spectrum.
In section 4,  we explain the nontrivial exponent of the distribution function at the critical point, through the correlation function matrix.
In section 5, we discuss the asymptotic form of the snapshot entropy.
In section 6, we present analysis of the 3-state Potts model.
The conclusion is summarized in section 7.

\section{Model and Snapshot Spectrum}

We consider the 2D ferromagnetic Ising model with the periodic boundary condition. 
The Hamiltonian is written as
\begin{eqnarray}
H=-J\sum_{y=1}^{N_y}\sum^{N_x}_{x=1} [S_{x,y}S_{x+1,y}+S_{x,y}S_{x,y+1} ],
\label{eq:ising_hamiltonian}
\end{eqnarray} 
where ${S_{x,y}=\pm 1}$ is the Ising spin variable and $x(y)$ denotes the lattice indices in the $x(y)$ direction.
In this paper, the exchange coupling is fixed at ${J=1}$. 
The linear dimensions in the $x$- and $y$-directions are respectively denoted as $N_x$ and $N_y$.
In the following, we assume $N_y \ge N_x$ and write the aspect ratio as $Q\equiv N_y/N_x$.

\begin{figure}[ht]
    \begin{center}
      \includegraphics[width=8cm]{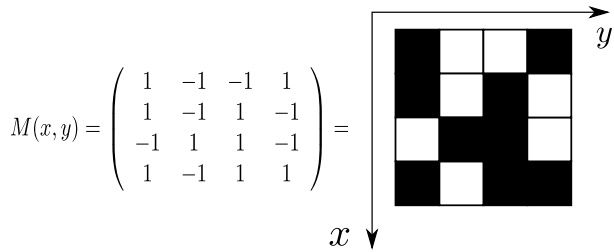}
      \caption{ Schematic of the snapshot matrix $M(x,y)$ for the Ising model on a $4\times 4 $ lattice. The solid or white squares indicate $S_{x,y} =\pm 1$. The orientation of the axis is adjusted to the matrix arrangement.
        }
      \label{fig1}
    \end{center}
\end{figure}

Suppose that a snapshot of the spin configuration at the equilibrium is generated by a MC simulation.
As in Fig. \ref{fig1}, we then regard this snapshot as an $N_x \times N_y$ matrix, the element of which is defined as $M(x,y)\equiv S_{x,y}$.
In the conventional MC simulation, the sample average of physical quantities is taken for a huge number of snapshots generated during a MC simulation.
In this paper, however, we directly consider SVD of the single snapshot matrix
\begin{eqnarray}
M(x,y)=\sum_{n=1}^{N_x}U_n(x)\Lambda_nV_n(y),
\label{eq:sc} 
\end{eqnarray}
where  $\Lambda_n $ is a singular value and $U_n $ and $ V_n$ denote the corresponding column vectors satisfying $U^t U = V^t V =1$.
Note that the number of singular values is $N_x$.

Since a snapshot is a  typical spin configuration at the equilibrium, it is expected that universal features associated with the phase transition can be extracted from the singular value spectrum $\Lambda$.
However, the singular value spectrum is practically difficult to treat in numerical computations.
We thus define a snapshot density matrix (SDM) as 
\begin{eqnarray}
\rho_X(x,x')=\frac{1}{N_y} \sum^{N_y}_{y}M(x,y)M(x',y), 
\label{eq:Nrhox}
\end{eqnarray}
or
\begin{eqnarray}
\rho_Y(y,y')=\frac{1}{N_x} \sum^{N_x}_{x}M(x,y)M(x,y'), 
\label{eq:Nrhoy}
\end{eqnarray} 
where we trace out the $y (x)$ component of $M(x,y)$.
In the case of the Ising model, the diagonal elements of $\rho_X (\rho_Y)$ are unity, which  naturally leads us to the normalization, $\Tr \rho_{X} = N_x$($ \Tr \rho_Y= N_y$).
These SDMs are real symmetric matrices, which are easy to handle with the conventional Householder diagonalization.

We consider the spectrum of $\rho_X$. 
Substituting Eq. (\ref{eq:sc}) into Eq. (\ref{eq:Nrhox}), we obtain
\begin{eqnarray}
\rho_X(x,x')=\frac{1}{N_y}\sum^{N_x}_{n=1}U_n(x)\Lambda_n^2U_n(x'),
\label{eq:rhoxx} 
\end{eqnarray} 
where the eigenvalues satisfy $\Lambda_n^2 \ge 0$ and  are assumed to be arranged in descending order. 
Here, we define the normalized eigenvalue spectrum as 
\begin{equation}
\omega_n \equiv \frac{1}{N_y} \Lambda_n^2 ,
\end{equation}
which satisfies the normalization
\begin{eqnarray}
\sum^{N_x}_{n=1}\omega_n=N_x.
\label{eq:normalization}
\end{eqnarray}
Using the numerical diagonalization of the SDM, we can thus investigate the snapshot spectrum $\{\omega_n\}$ and the corresponding unitary matrix $U$ in detail.
In particular, we mainly discuss the distribution function of snapshot eigenvalues (density of states),
\begin{equation}
p(\omega) = \frac{1}{N_x}\sum_n \delta(\omega -\omega_n),
\end{equation}
for the sufficiently large $N_x$.

\section{Eigenvalue Distribution for the Ising Model}

In the practical computation of snapshots, we use Wolff's cluster algorithm\cite{Wolff} near the critical temperature, while, in the high-temperature region, we use the Swendsen-Wang algorithm\cite{SW}. 
The relaxation steps to the equilibrium state are typically $N_x \times N_y$.
For a given snapshot configuration, we diagonalize a SDM to obtain a snapshot spectrum  $\{\omega_n\}$.
Then, we approximate $p(\omega)$ using the histogram of the eigenvalue distribution, where the typical width of the $\omega$ discretization is $\Delta\omega=0.016 - 0.08$.
Of course, the distribution function  within a single snapshot contains a large statistical fluctuation.
Thus, we typically take $5 \times 10^3 - 1 \times 10^5$ sample averages.
Here, we comment on error bars in the following figures of the distribution function.
In the scale of the figures, the error bar is basically negligible in the region $p(\omega) \gsim 5 \times 10^{-3}$, for which detailed analysis of the distribution function will be performed.
While, for $p(\omega) \lsim 5 \times 10^{-3}$, the  error bar becomes nonnegligible. However, it will not be shown in the figures to improve the clarity.

\subsection{Ising model on the square lattice $(N \times N)$}

\begin{figure}[t]
    \begin{center}
      \includegraphics[width=7cm]{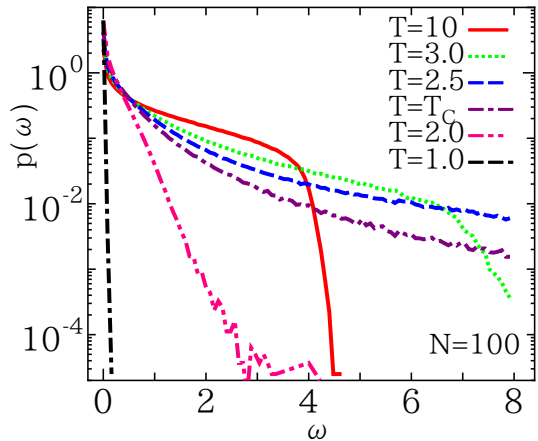}
      \caption{(Color online)
        Semilog plot of the eigenvalue distribution $p(\omega)$ for $N=100$ at $T=1.0 \sim 10.0$. 
        }
      \label{fig2}
    \end{center}
\end{figure}

We first consider the square-lattice Ising model of $N_x=N_y\equiv N$, for which both of the snapshot matrix and the SDM are $N\times N$ square matrices. 
In Fig. \ref{fig2}, we show a semilog plot of snapshot eigenvalue distributions for various temperatures $T$.
The features of the temperature dependence of the distribution functions are summarized as follows: 
(i) In the high-temperature phase, the eigenvalue distribution is in a finite range. 
(ii) In the low-temperature phase, the distribution has a very tall peak at $\omega=0$ and decays exponentially in the finite-$\omega$ region.
(iii) At the critical temperature, the snapshot spectrum becomes very broad and thus the distribution function exhibits a very slow decay in the large-$\omega$ regime.
Below, we analyze these characteristic behaviors of the  distribution function in detail, which are closely related to the nature of each phase.

Let us start with the high-temperature phase.
In the high-temperature limit,   spin configurations become random, where the thermal fluctuation is dominant and the correlation effect due to the energy is negligible.
If the snapshot matrix $M(x,y)$ is a random matrix, the corresponding ${\rho_{X}}$ is a Wishart matrix in RMT, the property of which is briefly summarized in Appendix.
Note that, for the square-lattice Ising model, the aspect ratio is $Q=1$ and the variance of the spin variable is $\sigma^2=1$, for which the lower and upper bounds of the eigenvalue distribution in $N\to \infty$ are respectively given by $\lambda_-=0$ and $\lambda_+=4$.

\begin{figure}[ht]
    \begin{center}
      \includegraphics[width=8.5cm]{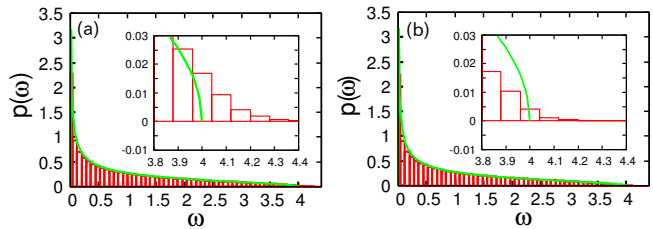}
      \caption{(Color online)
        Eigenvalue distribution $ p(\omega) $ for $N=100$ at (a) $ T=10.0$ and (b) $T=100.0$, where the discretization width of the histogram is $\Delta\omega=0.08$.
        We also plot the RMT distribution of $\sigma^2=1,Q=1$ for $N\to \infty$ (solid line).
        The inset is the enlarged view around the neighborhood of the maximum eigenvalue $\lambda_+$.
      }
      \label{fig3}
    \end{center}
\end{figure}

In Fig. \ref{fig3}, we show comparisons between the SDM eigenvalue distributions at $T=10$ and 100 with the corresponding RMT distribution (\ref{eq:RMT}) for $N\to \infty$ with $Q=1$ and $\sigma^2=1$.
In the figure, we can basically confirm the good agreement.
However,  it should be remarked that, in the principal component analysis of statistical data, the correlation effect generally emerges as eigenvalues beyond the RMT upper bound $\lambda_+$.
Thus, let us check the distributions around $\lambda_+$ precisely.
In the inset of Fig. \ref{fig3}(a),  we actually find that the eigenvalue distribution has a tail exceeding the upper bound $\lambda_+$, whereas, in the inset of Fig.\ref{fig3} (b), the eigenvalue distribution at $T=100$ basically falls within $\lambda_+$.
Thus, it is confirmed that the lager eigenvalues beyond $\lambda_+$ are induced by the spin correlation effect.
Here, we have checked that the distributions beyond $\lambda_+=4$ in Fig.\ref{fig3} are not a finite-size effect, with computations up to $N=280$.
As the temperature decreases further, the eigenvalue distribution extends to the larger-$\omega$ region, as depicted in Fig. \ref{fig2}.

We turn to the low-temperature phase, where dominant spins in a snapshot are aligned in the same direction.
Since matrix elements of $M$ belonging to the dominant percolating cluster are the same, the SDM becomes linearly dependent, so that the  matrix rank of the SDM effectively decreases.
In particular, at $T=0$, the maximum eigenvalue $\omega_1=N$ and the others are zero.
This implies that, in the low-temperature phase,  the huge maximum eigenvalue  $\omega_1 \propto N$ is solely located away from the main distribution, and the $\delta$-function-like peak appears at $\omega=0$, reflecting the condensation of the macroscopic number of zero eigenvalues.
Note that, in Fig. \ref{fig2}, the location of the maximum eigenvalue $\omega_1$ for $T=1.0$ and 2.0 is far out of the range of the horizontal axis.
The peak heights at $\omega=0$ for $T=1.0$ and 2.0 are also much larger than the range of the vertical axis.
The distribution of the remaining finite eigenvalues appears in a finite-$\omega$ region, where the exponential decay $p(\omega) \sim \exp(- {\rm const}\, \omega )$ is observed. 
Thus, we can characterize the ordered phase as the zero-eigenvalue condensation of the SDM spectrum.

\begin{figure}[ht]
    \begin{center}
      \includegraphics[width=7cm]{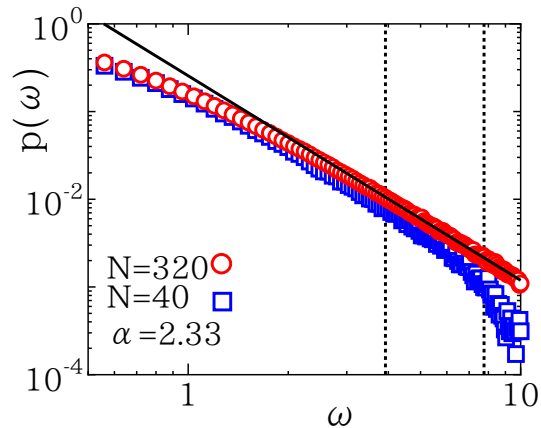}
      \caption{(Color online)
       Log-log plots of the eigenvalue distributions $p(\omega)$ at $T_c$ for $N=40$ and 320. 
The data for $N=40$ exhibits the finite-size effect in the large-$\omega$ region.
The solid line represents the power-law distribution (\ref{eq:pow}) with a nontrivial exponent $\alpha=2.33$,  which is estimated by the fitting for $N=320$.
The vertical dotted lines indicate the window  $\omega \in [4.08,7.12]$, which is used for the fitting.
      }
      \label{fig4}
    \end{center}
\end{figure}

As seen in Fig.  \ref{fig2}, the distribution at the critical temperature $T_c$ becomes very broad.
According to  the standard theory of the critical phenomenon,  $T_c$ is nothing but the percolation threshold, where snapshots involve spin clusters significantly fluctuating in the macroscale.
Then, it is naturally expected that the eigenvalue distribution at $T_c$ shows the power-law behavior, 
\begin{equation}
p(\omega) \propto \omega^{-\alpha}
\label{eq:pow}
\end{equation}
for $\omega\gg1$.
In Fig. \ref{fig4}, we present log-log plots of the distribution functions at $T_c$  for $N=40$ and $320$, where we can verify the linear behavior in a large-$\omega$ region.
For the region where the finite-size effect is negligible, we perform the fitting of  $p(\omega) \propto A \omega^{-\alpha}$.
If we adopt $\omega \in [4.08,7.12]$ as a fitting window, we obtain $\alpha \simeq 2.33$.
In Fig.\ref{fig4}, this fitting result is drawn as a solid line, which is consistent with the numerical result of $p(\omega)$.
Thus, we have confirmed that  the power-law distribution is actually realized at $T_c$.
We will discuss the theoretical background of this nontrivial power of $\alpha$ in Sect. \ref{sec:cfm}.

\subsection{Behavior of the high-ranking eigenvalues at $T_c$}
\label{sec:ising_eigen}

As seen above,  the power-law behavior of the distribution function is direct evidence of the critical behavior.
In determining the critical temperature, however, direct confirmation of such a power-law behavior is not very useful. 
Instead,  one can often use the finite-size-scaling analysis of Binder cumulant, which becomes independent of the system size at $T_c$.\cite{Binder}

\begin{figure}[ht]
    \begin{center}
      \includegraphics[width=8.5cm]{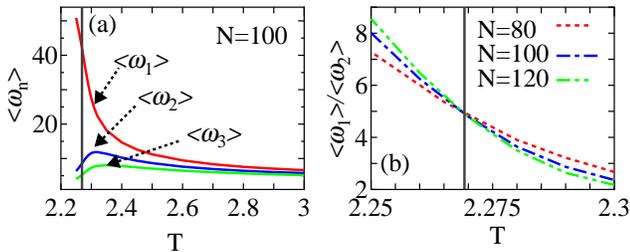}
      \caption{(Color online)
        (a) Temperature dependences of the high-ranking eigenvalues for $ N=100 $ in $T=2.2 \sim 3.0$.
The curves indicate $\langle\omega_1\rangle$, $\langle\omega_2\rangle$, and  $\langle\omega_3\rangle$ from top to bottom.
The black solid line represents the exact critical temperature $T_c$.
        (b) Temperature dependence of the ratio $\langle\omega_1\rangle/\langle\omega_2\rangle$ for $N=80 \sim 120$.
The curves cross at $T_c$, which is indicated as a vertical solid line.
        }
      \label{fig5}
    \end{center}
\end{figure}

Here, let us focus on the size dependence of the high-ranking eigenvalues.
In Fig. \ref{fig5} (a), we show the temperature dependence of the high-ranking eigenvalues $\omega_n$ for $n=1,2$, and $3$.
We take a 10000-sample average to obtain the curves in the figure, where  the error bar is not visible in the scale of the vertical axis.
As the temperature decreases from the high-temperature limit, the short-range correlation develops spin clusters, implying that the high-ranking eigenvalues increase gradually.
 As the temperature decreases below $T_c$, the maximum eigenvalue $\langle\omega_1\rangle$ rapidly increases, whereas $\langle\omega_2\rangle$ and $\langle\omega_3\rangle$ decrease.
These behaviors are consistent with the fact that the matrix rank of the SDM collapses toward unity in the zero-temperature limit, where $\langle\omega_1\rangle$ acquires the macroscopic scale and the other eigenvalues fall into zero.

In Fig.\ref{fig5} (a), the eigenvalues crossover  between the high- and low-temperature behaviors slightly above the critical point.
In order to determine the critical point, however, a careful analysis of the finite-size effect is needed.
Taking account of the splitting behavior of $\langle\omega_1\rangle$ and $\langle\omega_2\rangle$ below $T_c$, we examine the ratio $\langle\omega_1\rangle/\langle\omega_2\rangle$.
Figure \ref{fig5} (b) shows the temperature dependence of  $\langle\omega_1\rangle/\langle\omega_2\rangle$ for $N=80\sim 120$.
In the figure, the size dependences in the high- and low-temperature phases exhibit the opposite behaviors; moreover, the curves clearly cross at the exact $T_c$, which is indicated as a vertical solid line.
This suggests that the ratio $\langle\omega_1\rangle/\langle\omega_2\rangle$ becomes size-independent at $T_c$, like the Binder cumulant.
We can therefore determine the transition temperature using the ratio $\langle\omega_1\rangle/\langle\omega_2\rangle$.
The theoretical background of this behavior of the ratio $\langle\omega_1\rangle/\langle\omega_2\rangle$ will also be discussed in Sec. \ref{sec:cfm}.

\subsection{Ising model on the rectangular lattice $(N_x < N_y )$}

We next consider how the aspect ratio of the system affects the eigenvalue distribution, where the snapshot matrix $M(x,y)$ is rectangular.
The system size is taken to be $N_x \times N_y$ with $N_x\equiv N$ and  $N_y=2 N$.
An interesting point for the rectangular lattice is that the eigenvalue distribution in the  high-temperature limit is described by  RMT (Eq.(\ref{eq:RMT})) with $ Q\ne 1$.
Then, the range of the RMT distribution for $N\to \infty$ is different from that in the square case. 
In particular, we have $\lambda_- =0.0858$ for $Q=2$, which suggests that the eigenvalue distribution at $\omega=0$ is absent for $T>T_c$.

\begin{figure}[ht]
    \begin{center}
      \includegraphics[width=6cm]{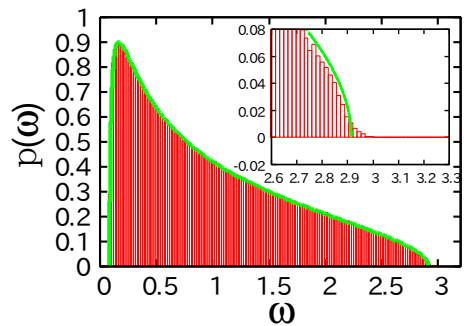}
      \caption{(Color online)
        Eigenvalue distributions $ p(\omega) $ at T=100 for $ N=200$, where $\Delta \omega=0.016$.
        We also plot the RMT curve for $N\to \infty$ with $\sigma^2=1$ and $ Q=2$ as a solid line.
        The inset shows the enlarged view around the upper bound of $\lambda_+$.}
      \label{fig6}
    \end{center}
\end{figure}

In Fig.\ref{fig6}, we show the eigenvalue distribution for $N=200$ at $T=100$.
In the figure, the eigenvalue distribution is basically in agreement with the  RMT curve, which is drawn as a solid curve.
In the inset, we can also see that the distribution beyond the RMT upper bound $\lambda_+$ is very small, implying that $T=100$ is a sufficiently high temperature.
Note that the finite-size effect was checked to be negligible for calculations up to $N=200$.

\begin{figure}[ht]
    \begin{center}
      \includegraphics[width=8.5cm]{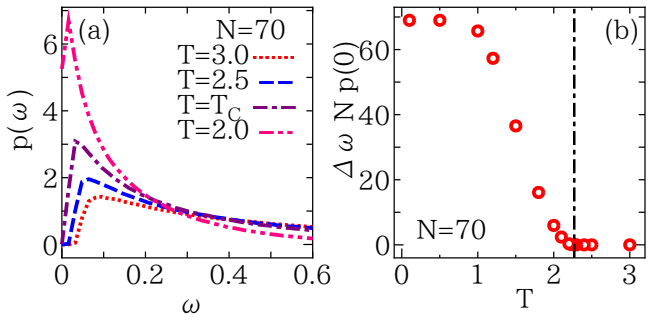}
      \caption{(Color online)
        (a) Temperature dependence of the eigenvalue distribution $ p(\omega) $ of $ N=70$ for T=2.0 $\sim$ 3.0. 
        (b) The peak height of the histogram at  $\omega=0$, where $\Delta\omega=0.016$. 
The zero-eigenvalue condensation occurs below $T_c$.
        }
      \label{fig7}
    \end{center}
\end{figure}

In Fig.\ref{fig7} (a), we show the temperature dependence of the eigenvalue distribution.
As the temperature decreases, the distribution develops into the large-$\omega$ region, which is consistent with the square-lattice case.
For the present rectangular lattice, another characteristic point is that the zero eigenvalue does not exist in the high-temperature phase.
We thus  plot the temperature dependence of the height of the histogram at $\omega=0$ in Fig. \ref{fig7} (b), which  illustrates that the zero eigenvalue appears only below $T_c$.
This behavior should be contrasted to the previous square-lattice result where a finite (but not macroscopic) number of eigenvalues appears at $\omega=0$ even in the high-temperature phase.
In this sense, the zero-eigenvalue condensation in the ordered phase can be easily verified for $Q \neq 1$.

\begin{figure}[ht]
    \begin{center}
      \includegraphics[width=6.5cm]{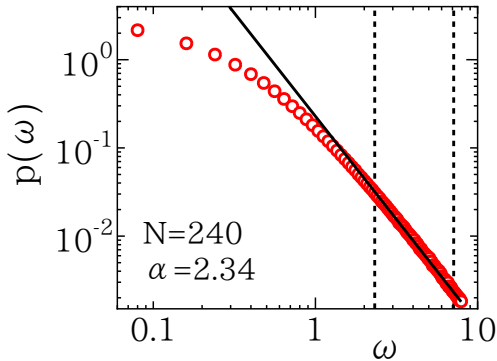}
      \caption{(Color online)
        Log-log plot of the eigenvalue distribution for $N=240$ at $T_c$.
        The solid line represents the fitting result based on the power-law distribution.
The vertical dotted lines indicate the window  $\omega \in [2.32,7.12]$, which is used for the fitting.
        }
      \label{fig8}
    \end{center}
\end{figure}

At the critical point,  the eigenvalue distribution for the rectangular lattice also exhibits the power-law behavior for $\omega\gg 1$.
Figure \ref{fig8} shows the log-log  plot of the eigenvalue distribution for $N=240$.
Note that the finite-size effect was confirmed to be negligible within the data plotted in the figure.
We perform  the same fitting as that in the square-lattice case  $[p(\omega) \propto A \omega^{-\alpha}]$, and plot its result as a solid line.
 The critical index is obtained as $\alpha \simeq 2.34$ for data in $\omega\in [2.32,7.12]$. 
Although the evaluation of $\alpha$ has a weak dependence on the range of a fitting window, the result is basically consistent with the square-lattice result.
Thus, the power-law distribution of $p(\omega)$ at $T_c$ is the universal feature independent of the shape of the lattice.

\section{Relation with the Correlation Function Matrix}
\label{sec:cfm}

In the previous section, we numerically found that the eigenvalue distribution of the SDM at $T_c$ obeys the power-law distribution $p(\omega) \propto \omega^{-\alpha}$ with $\alpha\sim 2.33$. 
In this section, let us consider the theoretical origin of this nontrivial power of $\alpha$.
A key point is that the matrix element of the SDM $\rho_X(x,x')= \sum_y M(x,y) M (x'y)$ can be regarded as an average of two spins separated by $x-x'$ with respect to the $y$-direction.
This average is of course  within a single snapshot, which involves the large statistical fluctuation.
In the thermodynamic limit, however, we can naturally expect $\rho_X(x-x') \sim G(x-x') \equiv  \langle S_{x,0} S_{x',0} \rangle$ by the self-averaging effect, where the translational invariance can be assumed for the periodic boundary system.
Thus, the asymptotic behavior of the SDM spectrum should be explained by the correlation function matrix $G(x-x')$, although the start point of our arguments is  at the SVD of the snapshot matrix $M$.

In general,  the correlation function at the critical point has the asymptotic form $G(x-x')\sim \left| x-x' \right|^{-d+2-\eta}$, where $\eta$ is the anomalous dimension.
In the thermodynamic limit, thus, each matrix element of the SDM with $d=2$ becomes
\begin{equation}
\rho_x(x,x') \sim G(x-x') \sim \left| x-x' \right|^{-\eta},
\label{rscfasymptotic}
\end{equation}
for $|x-x'| \gg 1$.
With help of the translational invariance, we can diagonalize the SDM by the Fourier transformation, $\omega(k)\sim |G(k)| \sim  |\sum_{r=0}^{N-1} r^{-\eta} \exp(ikr)|$, where $r=\left| x-x' \right|$ and $k= \frac{2 \pi n}{N}$.
Taking the limit $N \rightarrow \infty$, we evaluate $\omega(k)$ by the integral
\begin{equation}
\omega(k) \sim \left|\int_{0}^{\infty} r^{-\eta} \exp(i kr) dr \right| \sim |k^{\eta-1}| .
\end{equation}
The maximum eigenvalue is located at $k= 0$ with a certain long-distance cutoff.
The eigenvalues arranged in descending order correspond to  wave numbers $k=\frac{2 \pi n}{N}$ with $n=\pm 1, \pm 2, \cdots$.
Thus, the number of eigenstates bigger than a certain value $\omega$ can be counted as $n(\omega) \sim \omega ^{1/(\eta-1)}$.
As a result, we have the distribution function (density of states)
\begin{equation}
p(\omega) = \frac{d n( \omega )}{d\omega} \sim \omega^{-\alpha},
\label{eq:ppow}
\end{equation}
with
\begin{equation}
\alpha = \frac{2-\eta}{1-\eta}.
\label{eq:exactalpha}
\end{equation}
In the case of the Ising model, the anomalous dimension is $\eta=1/4$, which yields $\alpha =7/3 \simeq 2.33$.
This value is in good agreement  with the results obtained in the previous section.

We further consider the size dependence of $\omega_1/\omega_2$, on the basis of the relation with the correlation function matrix.
At a temperature slightly away from the critical temperature,  the correlation function is described  by the Ornstein-Zernike form
\begin{equation}
G(r)\propto r^{-(d-1)/2}e^{-r/\xi},
\end{equation}
where $\xi$ is the correlation length.
Thus, the eigenvalue spectrum of the SDM is also obtained by the Fourier transformation 
\begin{equation}
\omega(k) \sim G(k) \sim \frac{1}{(k\xi)^2+1}.
\end{equation}
The first and second eigenvalues respectively carry the wave numbers $k=0$ and $2\pi/N$.
As a result, the ratio $\omega_1/\omega_2$ becomes
\begin{equation}
\frac{\omega_1}{\omega_2} \sim 1+\left(\frac{2\pi}{N} \right)^2\xi^2.
\end{equation}
At the critical point, the scaling dimension of $(\xi/N)^2$ is zero.
Thus, the temperature dependences of the ratio ${\omega_1}/{\omega_2}$ for various system sizes cross at the criticality.

In a practical computation, the sample average of $p(\omega)$ is taken after diagonalization of the SDM, whereas the spectrum of the correlation function matrix $G$ is obtained by diagonalization after the sample average is taken.
The numerical results in the previous section suggest that the leading behavior of the SDM spectrum is consistent with the correlation function matrix by the self-averaging effect.

\section{Snapshot Entropy}

On the basis of the snapshot spectrum, we further discuss the snapshot entropy.
For the square-lattice Ising model, the normalized eigenvalue spectrum of the SDM is $\lambda_n \equiv \omega_n / N$, for which we may define the snapshot entropy as
\begin{eqnarray}
S_\chi\equiv -\sum^{\chi}_{n=1}\lambda_n\ln\lambda_n =-\sum^{\chi}_{n=1}\frac{\omega_n}{N}\ln\frac{\omega_n}{N},
\label{eq:vne}
\end{eqnarray}  
with a cutoff dimension $\chi \leq N$. 
The appearance of this entropy is reminiscent of the entanglement entropy.
What are its implications?
In analogy with the holographic principle\cite{Ryu}, Matsueda conjectured that the asymptotic behavior of Eq. (\ref{eq:vne}) could be  $S_\chi \sim \frac{1}{6} \log \chi+ $const for a small-$\chi$ region and $S_\chi \sim \log \chi+ $const for a large-$\chi$ region\cite{Matsueda}.
However, the theoretical justification for these asymptotic behaviors is missing in the context of statistical mechanics, so that a precise verification is  desired.

As discussed in Sec. 4, the asymptotic behavior of the snapshot spectrum is described by the correlation function matrix.
Thus, assuming the power-law distribution (\ref{eq:ppow}), we evaluate the leading behavior of the snapshot entropy as 
\begin{equation}
S_\chi \simeq \int_{\tilde{\omega}}^\infty p(\omega) \omega\log \left(\frac{\omega}{N} \right) d\omega \sim \int_{\tilde{\omega}}^\infty \omega^{1-\alpha} \log\omega d\omega .
\label{seeint}
\end{equation}
Here, the integral bound $\tilde{\omega}$ is attributed to the normalized cutoff dimension $\chi/N$ through
\begin{equation}
\frac{\chi}{N} = \int_{\tilde{\omega}}^\infty p(\omega) d\omega \sim \tilde{\omega}^{\frac{1}{\eta-1}},
\label{seecfd}
\end{equation}
which yields $\tilde{\omega} \sim \chi^{\eta-1}$.
We therefore have the leading asymptotic relation as
\begin{equation}
S_\chi = b \chi^{\eta}\log \frac{\chi}{a},
\label{seechi}
\end{equation}
where a cutoff scale  $a$ is recovered and $b$ is a certain overall coefficient.
Note that $a$ and $b$ have $N$ dependence for a finite-size system.

\begin{figure}[ht]
    \begin{center}
      \includegraphics[width=6.5cm]{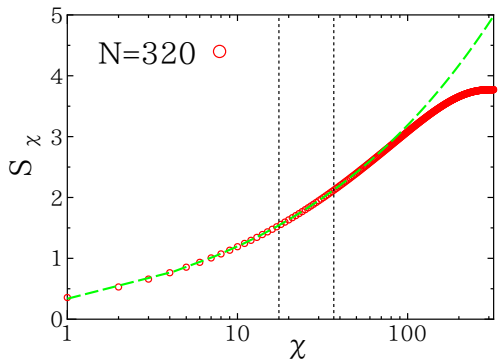}
      \caption{(Color online)
        Snapshot entropy $S_\chi$ for the square-lattice Ising model.
        The broken curve represents the asymptotic relation (\ref{seechi}) with the exact $\eta$.
        }
      \label{fig:seechi}
    \end{center}
\end{figure}

In order to confirm the theoretical prediction (\ref{seechi}), we compute the $\chi$ dependence of the snapshot entropy $S_\chi$ for $N=320$, where a 100-sample average is taken. 
The result is shown in Fig. \ref{fig:seechi}, where the error bar is not presented.
In the figure, we also plot the asymptotic relation $S_\chi = b \chi^{1/4}\log(\chi/a)$ with $a=0.1$ and $b=0.146$.
In comparison with the numerical result, it should be recalled that the relation (\ref{seechi}) is justified in the region where the power-law behavior of $p(\omega)$ is well established.
In Fig. \ref{fig4}, we adopted the region of $\omega \in [4.08,7.12]$ as an asymptotic regime.
The corresponding range of $\chi$ is $\chi \in [17,37]$, which is indicated by vertical dotted lines in Fig.\ref{fig:seechi}.
In this range of $\chi$,  we determine the parameters $a$ and $b$ so as to reproduce the numerical result well.
Then, the theoretical curve is in good agreement with the numerical result in a wide range beyond the original window of $\chi$.
We therefore verify that Eq. (\ref{seechi}) shows the correct asymptotic behavior of the snapshot entropy, rather than the naive logarithmic behavior proposed in Ref. [\citen{Matsueda}].
As $\chi$ approaches $N$, the theoretical curve deviates from the numerical result. 
This is because the correlation function is not described by the asymptotic form (\ref{rscfasymptotic}), where the short-range correlation is dominant.

\section{Eigenvalue Distribution of the 3-state Potts Model on the Square Lattice $(N \times N)$}

In order to check the universality of the above-mentioned results,  let us examine the 3-state Potts model,\cite{Potts}
\begin{eqnarray}
H=-J\sum^{}_{x,y} [ \delta(S_{x,y},S_{x+1,y}) + \delta(S_{x,y},S_{x,y+1})] ,
\label{eq:potts_hamiltonian}
\end{eqnarray} 
where  ${S=\pm 1,0}$.
Also, ${J(=1)}$ denotes the coupling constant, and $\delta(S,S')$ is the Kronecker's delta symbol.
The system size is taken to be $N_x = N_y = N$ and the periodic boundary condition is imposed.
Note that the exact critical temperature of the 3-state Potts model is ${T_c=0.99497\cdots}$.

The analysis of the Potts model is almost parallel to the Ising model.
We use the Wolff algorithm to generate snapshots.
For the Potts model, however, it should be noted that the snapshot matrix (\ref{eq:sc}) depends on the definition of the spin variable;  for instance, $S = 0, \pm 1$ is not a $Z_3$ invariant variable of the spin.
In the following,  we thus define the snapshot matrix as
\begin{equation}
M(x,y)= \delta(S_{x,y},q) -\frac{1}{3},
\label{pottssm}
\end{equation}
where $q$ is fixed at one of $0,\pm 1$. 
This matrix can be regarded as a spin-resolved snapshot matrix and is independent of the definition of the spin variable.\cite{pottsnote}
The SDM for the Potts model is also defined as Eq. (\ref{eq:Nrhox}) with Eq. (\ref{pottssm}).
Here, it should be remarked that the normalization  $\Tr \rho_X = \sum \omega_n \simeq 2N_x/9$ is satisfied within the level of the average in the disordered phase, in contrast to the Ising model where Eq. (\ref{eq:normalization}) is always exact.
In calculating the eigenvalue distribution, the typical number of samples is up to $ 5 \times 10^3$.

\begin{figure}[ht]
    \begin{center}
      \includegraphics[width=6.5cm]{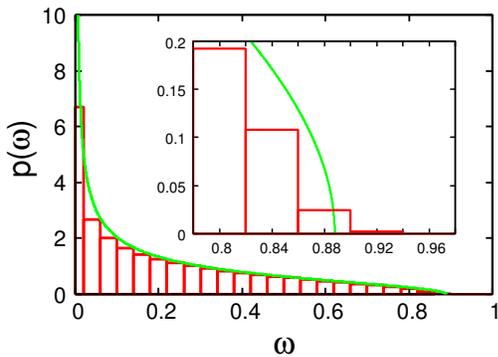}
      \caption{(Color online)
        Eigenvalue distribution $ p(\omega) $ for the 3-state Potts model of system size $N=100$  at $T=100.0$, where $\Delta\omega=0.04$.
        We also plot the RMT curve for $N\to \infty$ with  $\sigma^2=2/9$ and $Q=1$ as a solid line.
        The inset is an enlarged view around the upper bound $\lambda_+=8/9$ of the RMT curve.
        }
      \label{fig10}
    \end{center}
\end{figure}

In Fig. \ref{fig10}, we show the eigenvalue distribution of the SDM for the square-lattice 3-state Potts model of $N=100$ at $T=100$ with  $\Delta\omega=0.04$.
Here, note that the mean and variance of $M(x,y)$ for a random spin variable $S$ are respectively evaluated as $\langle \delta(S,q)-\frac{1}{3} \rangle =0$ and $\langle (\delta(S,q)-\frac{1}{3})^2 \rangle=\frac{2}{9}$.
In the figure, the RMT curve of Eq.(\ref{eq:RMT}) corresponding to $\sigma^2 = 2/9$ and $Q=1$ is illustrated as well.
We can verify the good agreement between the simulation result and the RMT curve.
Thus, the RMT description of the eigenvalue distribution is also valid for the 3-state Potts model.

\begin{figure}[ht]
    \begin{center}
      \includegraphics[width=6.5cm]{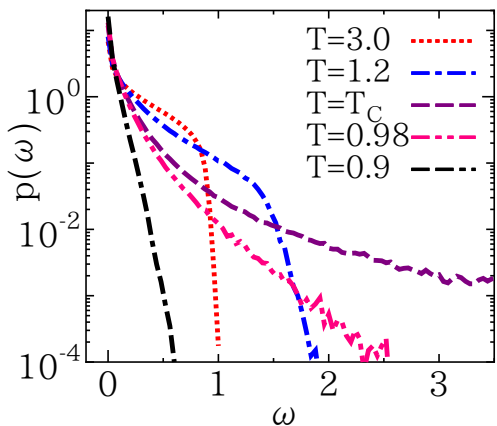}
      \caption{(Color online)
Semilog plot of the eigenvalue distribution $p(\omega)$ for $N=100$ at T=3.0$ \sim $0.9.
        }
      \label{fig11}
    \end{center}
\end{figure}

In Fig. \ref{fig11}, we show the temperature dependence of the distribution function $p(\omega)$ for $N=100$ and $T=0.9\sim 3.0$, where we can basically see behaviors  similar to the Ising model.
As the temperature decreases,  the cluster containing the same spin becomes larger, so that the distribution function develops beyond the bound $\lambda_+$.
At $T=T_c$, $p(\omega)$ shows the long-tail behavior in the large-$\omega$ region.
For $T<T_c$, moreover,  the zero-eigenvalue condensation clearly occurs, where the histogram at $\omega=0$ is on the order of $N$.
Here, it should be noted  that  the $Z_3$ symmetry is broken in the ordered phase, so that there appear two types of snapshot spectrum depending on whether or not the ordered spin coincides with  $q$.
Figure \ref{fig11} shows the distribution for the case where $q$ corresponds to the ordered spin.

\begin{figure}[ht]
    \begin{center}
      \includegraphics[width=8.5cm]{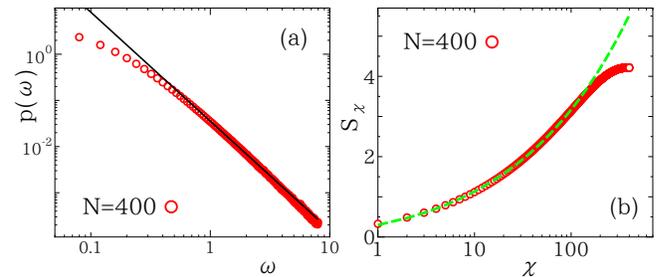}
      \caption{(Color online)
        (a) Log-log plot of the eigenvalue distribution for $N=400$ at $T_c$.        We also plot the guide line of the power-law distribution with the exact exponent $\alpha=2.3636\cdots$.
        (b) $\chi$ dependence of the snapshot entropy $S_\chi$ for the 3-state Potts model. 
The broken line indicates the asymptotic curve of Eq. (\ref{seechi}) with the exact $\eta$.
        }
      \label{fig12}
    \end{center}
\end{figure}

At the critical temperature, we can expect that $p(\omega)$ obeys the power-law distribution.
In Fig.\ref{fig12} (a), we show the log-log plot of the eigenvalue distribution for $N=400$ at $T_c$.
In the case of the 3-state Potts model, the exact value of the anomalous dimension is  $\eta=4/15$, which yields $\alpha =26/11= 2.3636\cdots$. 
In Fig. \ref{fig12} (a), we also draw the slope with the exact $\alpha$ as a solid line.
Although $p(\omega)$ still contains a small finite-size effect even for $N=400$,  we consider that the numerical result is basically consistent with Eq. (\ref{eq:exactalpha}).

We finally discuss the $\chi$ dependence of the snapshot entropy for the 3-state Potts model.
In Fig. \ref{fig12} (b), a comparison between the simulation result and the asymptotic curve of Eq. (\ref{seechi}) with the exact $\eta$ is presented.
The good agreement can be confirmed in the wide range of $\chi$, as in the case of the Ising model.
We therefore concluded that the snapshot entropy at the critical point is also described by the theory based on the correlation function matrix.

\section{Summary and Discussion}
In summary, we have investigated the eigenvalue distributions of SDMs generated by MC simulations for the 2D Ising and 3-state Potts models.
We have found that the eigenvalue distribution captures the essential features of the phase transition.
The high-temperature limit is described by the  Wishart-type RMT, whereas the low-temperature phase is characterized by the zero-eigenvalue condensation, which is attributed to the appearance of  the percolation cluster below $T_c$.
 We also find that the eigenvalue distribution of the SDM at $T_c$ obeys the power-law distribution $p(\omega) \propto \omega^{-\alpha}$.
The relation with the correlation function matrix enables us to derive the analytic formula of the nontrivial power $\alpha=(2-\eta)/(1-\eta)$, which is consistent with the numerical results.
We have also derived the asymptotic form of the $\chi$ dependence of the snapshot entropy $S_\chi\sim \chi^\eta \log(\chi/a)$.
Since this relation  successfully explains the numerical result in a wide range of $\chi$,  we think that it is a correct asymptotic form of $S_\chi$ rather than the naive logarithmic dependence proposed in Ref. [\citen{Matsueda}].

The snapshot spectrum may be a different concept from the entanglement spectrum in the quantum system, although our motivation originally came from the entanglement for the quantum system. 
As shown in this paper, however, the snapshot spectrum can be used to extract the essential features of the phase transition.
Moreover, the snapshot is easy to handle in the MC simulation, in contrast to a direct treatment of the maximal eigenvector of the transfer matrix.
We thus consider that the present analysis provides further perspectives in analyzing the phase transitions of various spin systems.
In addition, how the relations (\ref{eq:ppow}) and (\ref{seechi}) can be associated with  the quantum many-body system is an important problem.
Then, the fact that the correlation function matrix has a direct connection with the entanglement Hamiltonian for the free fermion system\cite{Peschel}  may provide an interesting hint to address this problem.

\section*{Acknowledgements}
One of the authors (K.O) would like to thank H. Matsueda for valuable discussions.
This work was supported by Grants-in-Aid Nos. 23540442 and 23340109 from the Ministry of Education, Culture, Sports, Science and Technology of Japan. 
It was also supported in part by the Strategic Programs for
Innovative Research (SPIRE), MEXT, and the Computational Materials Science Initiative (CMSI), Japan.


\appendix

\section{Random Matrix Theory}
In this appendix, we briefly summarize the eigenvalue distribution of the Wishart random matrix.\cite{RMT}
Write a matrix of $N \times L$ as $A$, the element of which is independently defined by a random real number of zero mean and  variance $\sigma^2$.
Note that $\sigma = 1$ for the Ising variable.
Then, the eigenvalue distribution of the Wishart-type matrix $X\equiv \frac{1}{L} A A^t$ in $N,L\to \infty$ with ${Q \equiv L/N}(={\rm const.})$ is given by
\begin{eqnarray}
p(\lambda)=\frac{Q}{2\pi \sigma^2}\frac{\sqrt[]{(\lambda_+-\lambda)(\lambda-\lambda_-)}}{\lambda}.
\label{eq:RMT}
\end{eqnarray}
Here,  $\lambda_-$ and $\lambda_+$ respectively denote the lower and upper bounds of the eigenvalue spectrum, which are explicitly given by 
\begin{eqnarray}
\lambda_\pm=\sigma^2\left(1+\frac{1}{Q}\pm 2\sqrt[]{\frac{1}{Q}}\right).
\label{eq:RMT_lam}
\end{eqnarray}
Note that, for $Q=1$, $\lambda_-=0$ and thus $p(\lambda)$ diverges as $p(\lambda) \sim \lambda^{-1/2}$ in $\lambda \to 0$.
However, this divergence does not indicate the zero-eigenvalue condensation on the macroscopic scale.


\begin{thebibliography}{99}
\bibitem{pottsnote} In Ref. [\citen{Matsueda}], the snapshot matrix is directly constructed with the spin variable $S=0,\pm1$.
This definition of the snapshot matrix gives an equivalent spectrum to the definition of Eq. (\ref{pottssm}) except for the normalizations.
\end{thebibliography}
\end{document}